\documentclass{article}

\usepackage{conference,times}

\usepackage[utf8]{inputenc}
\usepackage[T1]{fontenc}
\usepackage{amsmath,amssymb}
\usepackage{graphicx}
\usepackage{booktabs}
\usepackage{multirow}
\usepackage{subcaption}
\usepackage{hyperref}
\usepackage{natbib}
\usepackage{tikz}
\usetikzlibrary{arrows.meta,positioning,fit,backgrounds}
\usepackage{float}

\iclrfinalcopy

\title{Correcting the Dropout-LayerNorm \\ Expectation Gap Improves Protein Structure Models}

\author{Isaac Ellmen$^1$, David Errington$^2$, Matthew I.J. Raybould$^1$, Charlotte M. Deane$^1$\thanks{Corresponding author: \texttt{deane@stats.ox.ac.uk}} \\
$^1$Department of Statistics, University of Oxford \\
$^2$Recursion
}

\begin{document}

\maketitle

\begin{abstract}
Although $\mathbb{E}[\mathrm{dropout}(x)] = x$, here we show that $\mathbb{E}[\mathrm{LayerNorm}(\mathrm{dropout}(x))]$ is \emph{not} equal to $\mathrm{LayerNorm}(x)$.
Accordingly, the pattern of a Dropout layer followed by a LayerNorm, which is common to many AlphaFold2-based protein structure predictors, produces a systematic bias at evaluation time that can hamper performance.
To address this, we derive a closed-form, first-order correction for this gap, which we call a Dropout-LayerNorm Correction (DLC).
DLC empirically matches the performance boost of large Monte Carlo dropout ensembles.
We evaluate its effect across nine protein structure models (ESMFold, OpenFold, ABB3, FlashABB, Ibex, NbForge, Genie1, Genie2, Genie3) on both paired and single-chain antibody structures as well as one protein-ligand docking model (QuickBind).
The correction is computationally negligible and improves accuracy in all ten models tested ($\sim0.3\%-13\%$), with a modest but consistent improvement in ESMFold and OpenFold and a substantial improvement in antibody-specific models.
This work identifies the mathematical consequence of chaining together Dropout and LayerNorm and provides a free, principled adjustment to improve the evaluation performance of many pretrained models.
Code to reproduce the experiments is available at \url{https://github.com/oxpig/DLC}.
\end{abstract}

\section{Introduction}

Most modern neural networks are composed of a relatively small number of unique layers.
Among these are dropout and normalization layers, which are used to improve generalization and stability.
During training, dropout layers randomly replace each entry with a zero with probability $p$ \citep{srivastava_dropout_2014}.
This creates an imbalance during evaluation, since the magnitude of each entry will be, in expectation, $1/(1-p)$ larger.
To compensate for this, standard dropout implementations, such as the PyTorch \texttt{nn.Dropout} layer, multiply by $1/(1-p)$ during training.
This means that $\mathbb{E}[\mathrm{dropout}(x)] = x$ and empirically fixes the eval-mode gap \citep{srivastava_dropout_2014}.
However, underlying this fix is the assumption that dropout layers are always followed by a layer which is invariant to changes in variance, such as random linear layers.
In practice, sometimes dropout layers are followed by non-linear blocks such as residual+LayerNorm or, in the case of AlphaFold2, just LayerNorm \citep{ba_layer_2016, vaswani_attention_2017, dauparas_robust_2022, jumper_highly_2021}.

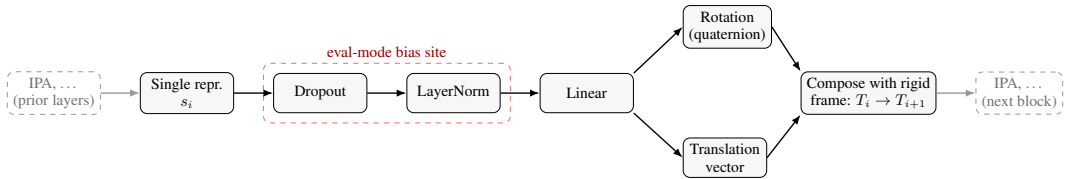
\begin{figure}[h]
\centering
\resizebox{\textwidth}{!}{%
\begin{tikzpicture}[
  node distance=6mm and 8mm,
  box/.style={draw, rounded corners, minimum height=8mm, minimum width=19mm,
              align=center, font=\small, fill=black!3},
  small/.style={draw, rounded corners, minimum height=7mm, minimum width=17mm,
                align=center, font=\footnotesize, fill=black!3},
  faint/.style={draw=black!40, rounded corners, minimum height=7mm,
                minimum width=17mm, align=center, font=\footnotesize,
                text=black!55, dashed},
  arr/.style={-{Latex[length=2mm]}, thick},
  farr/.style={-{Latex[length=2mm]}, thick, draw=black!40},
]
\node[box] (s) {Single repr.\\$s_i$};
\node[faint, left=of s] (prior) {IPA, \ldots\\(prior layers)};
\node[box, right=of s] (drop) {Dropout};
\node[box, right=of drop] (ln) {LayerNorm};
\node[box, right=of ln] (lin) {Linear};
\node[small, above right=5mm and 10mm of lin] (rot) {Rotation\\(quaternion)};
\node[small, below right=5mm and 10mm of lin] (trans) {Translation\\vector};
\node[box, right=34mm of lin] (compose) {Compose with rigid\\frame: $T_i \to T_{i+1}$};
\node[faint, right=of compose] (nextblock) {IPA, \ldots\\(next block)};

\draw[farr] (prior) -- (s);
\draw[arr] (s) -- (drop);
\draw[arr] (drop) -- (ln);
\draw[arr] (ln) -- (lin);
\draw[arr] (lin.north east) -- (rot.west);
\draw[arr] (lin.south east) -- (trans.west);
\draw[arr] (rot.east) -- (compose.north west);
\draw[arr] (trans.east) -- (compose.south west);
\draw[farr] (compose) -- (nextblock);

\begin{scope}[on background layer]
\node[draw=red!60, dashed, rounded corners, fit=(drop)(ln), inner sep=2mm,
      label={[red!60!black,font=\footnotesize]above:eval-mode bias site}] {};
\end{scope}
\end{tikzpicture}%
}
\caption{The dropout$\to$LayerNorm$\to$Linear$\to$rigid-body-update
pattern repeated at the end of every structure-module block.
The single representation $s_i$ passes through dropout and LayerNorm (the eval-mode bias site), then a Linear layer that directly outputs the block's rotation and translation update.
As the biased LayerNorm output feeds the Linear layer that \emph{is} the translation vector, the bias shows up directly as a systematic error in predicted
translation length.
}
\label{fig:block-diagram}
\end{figure}

The final layer of each StructureModule block in AlphaFold2-style models finishes with a Transition into a BackboneUpdate \citep{jumper_highly_2021, ahdritz_openfold_2024}.
Practically, this reads as Dropout$\to$LayerNorm$\to$Linear$\to$Translation (see Figure \ref{fig:block-diagram}).
This means that any bias in the expectation of dropout+LayerNorm is propagated directly into the expectation of the translation of each amino acid.
In the following sections, we show that this leads to a systematic ``overshoot'' of approximately $5-6\%$ which materially affects the quality of predicted structures.
However, we also provide a fix for this effect without the need for retraining.
We derive a closed-form correction which scales roughly as $\sqrt{q} = \sqrt{1-p}$.
This correction can be applied cheaply after every dropout+LayerNorm block and nearly perfectly fixes the overshoot and improves structural accuracy.

Our contributions are the following:

\begin{enumerate}
    \item We identify a systematic bias that exists in all neural networks with a dropout layer directly followed by a LayerNorm
    \item We provide a closed form fix which can be patched directly into LayerNorm
    \item We test our fix on ten biomolecular structure models in a therapeutically relevant protein context (immunoglobulins) and show that it improves accuracy in almost all cases
\end{enumerate}

\section{Theory}
\label{sec:theory}

\begin{figure}[h]
\centering
\includegraphics[width=\textwidth]{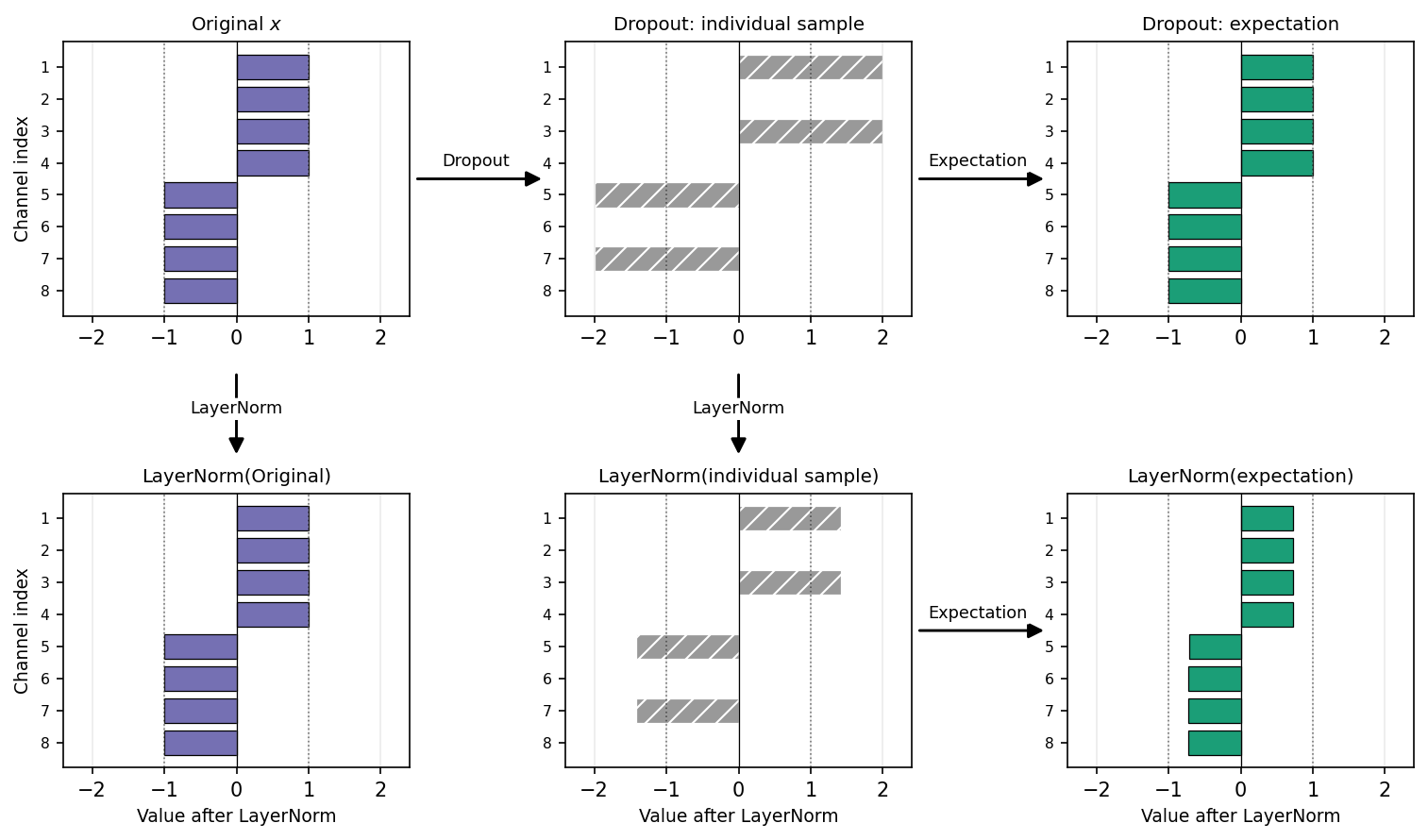}
\caption{Toy example isolating the mechanism ($d{=}8$, $q{=}0.5$,
$x=(1,1,1,1,-1,-1,-1,-1)$, so $\mu{=}0,\sigma{=}1$ by construction).
Dropout's own expectation exactly recovers $x$ (unbiased per channel,
middle-top), and LayerNorm alone is a no-op on this particular $x$
(bottom-left), but LayerNorm's expectation after dropout does not
recover $x$: it shrinks to $\sim\sqrt{q}\cdot x$ (bottom-right).
}
\label{fig:toy-example}
\end{figure}

\subsection{Setup}
Consider a single token's pre-dropout representation $x \in
\mathbb{R}^d$ immediately upstream of a LayerNorm.
Inverted dropout keeps each channel independently with probability $q$ (drop probability $p = 1-q$) and rescales by $1/q$:
\begin{equation}
z_k = \frac{m_k x_k}{q}, \qquad m_k \sim \mathrm{Bernoulli}(q) \text{ i.i.d.}
\end{equation}
This is exactly unbiased per channel ($\mathbb{E}[z_k] = x_k$) \citep{srivastava_dropout_2014}.
LayerNorm then computes
\begin{equation}
\mathrm{LN}(z)_k = \gamma_k \cdot \frac{z_k - \mathrm{mean}(z)}{\mathrm{std}(z)} + \beta_k ,
\end{equation}
where $\mathrm{mean}(z)$ and $\mathrm{std}(z)$ are shared across all $d$
channels.
This sharing is the entire source of the eval-mode bias.
In most cases, $\mathrm{mean}(z) \approx 0$, and so $\mathrm{LN}(z)_k$ is directly rescaled by changes in $\mathrm{std}(z)$, such as the variance shift induced by dropout.

\subsection{Simplified derivation (zero-mean case)}
\label{sec:simple-derivation}
We give the cleanest version of the result here, under the simplifying
assumption that the channel-wise mean of $x$ is (approximately) zero,
$\mu := \mathrm{mean}_k(x_k) \approx 0$.
For the non-zero mean derivation, see Appendix \ref{app:general-derivation}.

With $\mu \approx 0$, $\mathbb{E}[\mathrm{mean}(z)] = \mu \approx 0$, and
the shared-variance expectation
\begin{equation}
\mathbb{E}[\mathrm{var}(z)] \;\approx\; \frac{\sigma^2 + p\mu^2}{q}
\;=\; \frac{\sigma^2}{q},
\end{equation}
where $\sigma^2 := \mathrm{var}_k(x_k)$. Substituting into
$\mathbb{E}[\mathrm{norm}(z)_k] \approx (x_k - \mu)/\sqrt{\mathbb{E}[\mathrm{var}(z)]}$
gives
\begin{equation}
\mathbb{E}[\mathrm{norm}(z)_k] \;\approx\; \frac{x_k}{\sigma/\sqrt{q}}
\;=\; \sqrt{q}\cdot \mathrm{norm}(x)_k .
\end{equation}
So in the zero-mean case the correction collapses to a single global
constant, $c = \sqrt{q}$, independent of $\mu,\sigma$ altogether. Writing
$\mathrm{LN}_{\mathrm{full}}(x) := \gamma \cdot \mathrm{norm}(x) + \beta$
for LayerNorm's actual output (affine parameters included), the
$\beta$-preserving eval-time correction is
\begin{equation}
\label{eq:simple-correction}
\widehat{\mathrm{LN}}(x) \;=\; \sqrt{q}\cdot \mathrm{LN}_{\mathrm{full}}(x) \;+\; \left(1-\sqrt{q}\right)\beta .
\end{equation}
Only the $\gamma\cdot\mathrm{norm}(x)$ part is shrunk; $\beta$ is
recovered at full strength, which is exact algebraically (expand
Eq.~\ref{eq:simple-correction} and the $\beta$ terms collapse to
coefficient 1).

\subsection{General ($\mu \neq 0$) correction}

For non-zero $\mu$, the correction factor $c$ is well-approximated by $\sqrt{q/(1+p(\mu/\sigma)^2)}$.
This is the version we use throughout the remainder of the manuscript.
For a full derivation, see Appendix \ref{app:general-derivation}.
Intuitively, as the mean deviates from zero, the average effect of dropout becomes more extreme which increases the variance (and the required correction).
As the $\mu \to 0$, $c \to \sqrt{q}$ which recovers the simplified case.

In Appendix \ref{app:damped-derivation}, we also derive a correction for the case of $\mathrm{LN}(\mathrm{dropout(x')} + x)$, (i.e. a dropout vector added to the residual stream and then normalized).
Here, $c=\sqrt{q/(q+p\rho^2)}$ where $\rho$ is the ratio of the variance between $x'$ and $x$.
This is a common pattern in many neural networks such as GNNs and Transformers, but we found that practically, the effect was small enough to have limited impact on the models we tested.
We leave further exploration as an interesting direction for future work.
In Appendix \ref{app:other-norms}, we derive the equivalent correction for RMSNorm \citep{zhang_root_2019} and BatchNorm \citep{ioffe_batch_2015}, two other widely used normalization schemes which are not tested empirically in this paper.

\section{Methods}

We test our correction on protein structure predictors which contain the Dropout$\to$LayerNorm$\to$Linear$\to$Translation pattern.
This includes general protein structure predictors like OpenFold \citep{ahdritz_openfold_2024} and ESMFold \citep{lin_evolutionary-scale_2023} as well as a number of antibody-specific variants which omit the pair representation.
One of the first antibody-specific AlphaFold2-inspired antibody structure predictors was ImmuneBuilder \citep{abanades_immunebuilder_2023} which uses an analog of Invariant Point Attention with no pair processing.
Several similar models have since emerged including ABodyBuilder3 (ABB3) \citep{kenlay_abodybuilder3_2024}, FlashABB \citep{ellmen_modelling_2026}, and Ibex \citep{dreyer_conformation-aware_2025} for antibody structure prediction and NbForge \citep{ali_disulphide_2026} for nanobody/VHH structure prediction.
Similarly, Genie1 \citep{lin_generating_2023}, Genie2 \citep{lin_out_2024}, and Genie3 \citep{lin_fast_2026} use a variant of the OpenFold structure module for general protein structure generation.

Perhaps notably, many variants of OpenFold contain changes which avoid the phenomenon discussed in this work.
For instance, ImmuneBuilder uses the default 10\% dropout during the first stage of training but then uses 0\% dropout during the finetuning stage.
Similarly, FrameFlow uses a dropout rate of 0\% throughout the StructureModule \citep{yim_improved_2024}.
AlphaFold3-style models contain residual connections in the updates which may help to mitigate the variance shift \citep{abramson_accurate_2024, wohlwend_boltz-1_2024}.

We restrict our experiments to the study of antibodies and nanobodies.
This ensures a fair comparison between the models and allows us to use the established practice of measuring full-structure and Complementarity-Determining Region H3 (CDRH3)-specific RMSDs; CDRH3 is the most challenging region of an immunoglobulin to model.
For each model, we create a hook which fires post-LayerNorm at every site which is directly preceded by a dropout layer.
In most models this happens twice per block: once in the IPA transition and once in the backbone transition.
In Ibex, there is an additional hook which fires after the ESM-C language model embeddings \citep{candido_language_2026, dreyer_conformation-aware_2025}.
In NbForge, there is an additional hook for the pair updates \citep{ali_disulphide_2026}.

\section{Experiments}
\label{sec:experiments}

\subsection{Translation length is systematically wrong, and the correction fixes it}
\label{sec:block0}

\begin{figure}[h]
  \centering
  \includegraphics[width=\textwidth]{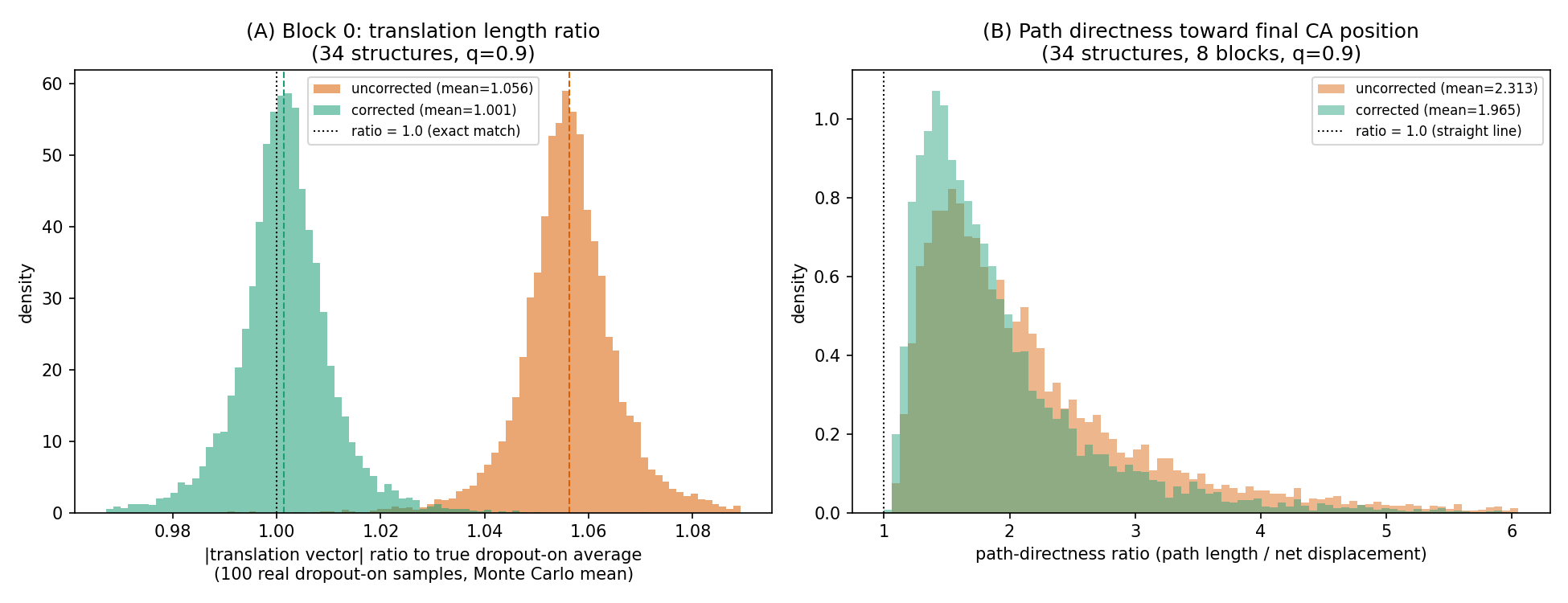}
  \caption{ABB3, both-sites corrected vs.\ uncorrected, 34-structure test
  subset, $q=0.9$. \textbf{(A)}: ratio of the first structure-module
  block's predicted backbone-translation vector length to the true
  100-sample Monte Carlo dropout-on average (mean $1.056\to1.001$).
  \textbf{(B)}: path-directness ratio (aggregate per-block C$\alpha$ displacement across all 8 blocks divided by net displacement from block 0 to block 7).
  Here, 1.0 corresponds to a perfectly straight path whereas higher values are associated with more ``zigzagging''.
  The corrected mean drops from 2.313 to 1.965, showing the block-0 magnitude bias is not an isolated artifact but part of a systematically less direct trajectory toward the final structure.}
  \label{fig:abb3-block0}
\end{figure}

We first sought to demonstrate that the dropout+LayerNorm eval gap does actually produce the predicted overshoot.
We performed an MC simulation of 100 random seeds with dropout on and measured the ratio of the length of the first translation (block 0) between the deterministic models (corrected + uncorrected) and the MC average for ABB3.
The results are shown in Figure \ref{fig:abb3-block0}.
We found that the uncorrected version did tend to overshoot the MC average by about 5.6\% ($\sim1/\sqrt{0.9}$) whereas our corrected version nearly perfectly matched the mean translation.
We also measured the ratio of the distance traveled to the total displacement over all residues for the corrected and uncorrected models.
The uncorrected model had to travel about 18\% further, consistent with a trajectory of chronically overshooting and needing to correct at every block.

\subsection{Theory matches the true Monte Carlo expectation}
\label{sec:mc-convergence}

\begin{figure}[h]
  \centering
  \includegraphics[width=\textwidth]{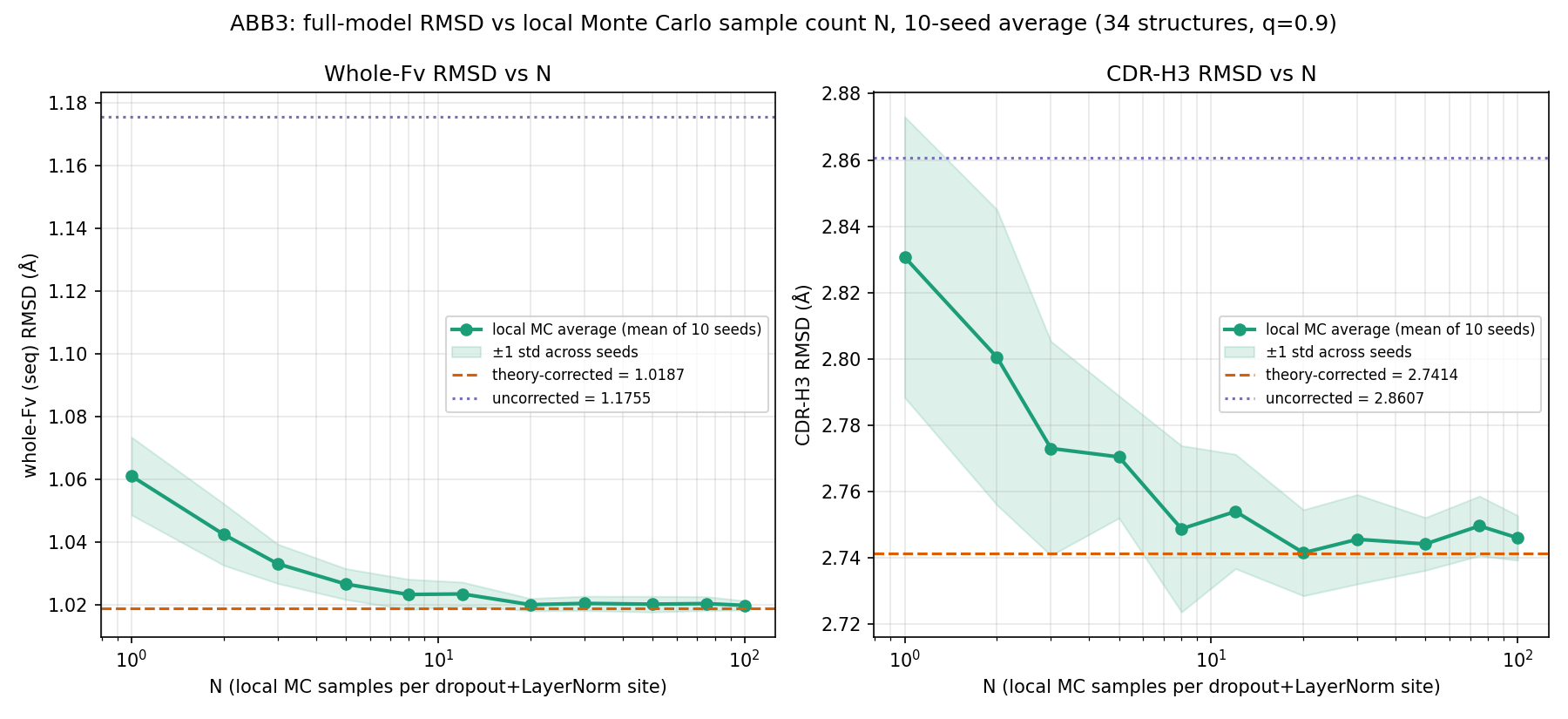}
  \caption{ABB3-predicted whole-Fv and CDR-H3 test RMSD as a function of
  $N$, the number of Monte Carlo samples used at every
  dropout+LayerNorm site (log-$x$), averaged over 10 independent seeds
  (shaded band: $\pm 1$ std across seeds). From $N{=}1$ the mean already
  sits well below the uncorrected line since we are introducing noise but removing bias.
  Both the mean and the across-seed spread converge onto the theory-corrected line by $N\!\approx\!30$, showing that the theory corrects to an unbiased approximation of the MC RMSD.
  }
  \label{fig:mc-convergence}
\end{figure}

We tested the impact of the correction on the final RMSD error of ABB3 structure predictions.
This test broadly matches similar works in MC dropout \citep{srivastava_dropout_2014, gal_dropout_2016, li_understanding_2018} where the mean prediction (coordinates) is taken over $n$ distinct samples.
The results are shown in Figure \ref{fig:mc-convergence}.
When measured over the entire structure and over only the CDRH3, the single-sample dropout-on evaluation was more accurate than the (standard) dropout-off evaluation.
Taking the mean coordinates improved accuracy monotonically with the number of samples.
This agrees with a model where dropout introduces an extra noise term which decays as $O(\frac{1}{\sqrt{n}})$.
The MC simulation also approached a structure with a similar RMSD to our corrected structure which shows that our correction maps well to the unbiased dropout-on mean.
Thus, our correction is a cheap approximation of the MC mean which also avoids complications with nonphysical structures induced by averaging coordinates.

\subsection{Expanding the Pareto front of antibody structure prediction}
\label{sec:headline}

\begin{figure}[h]
  \centering
  \includegraphics[width=\textwidth]{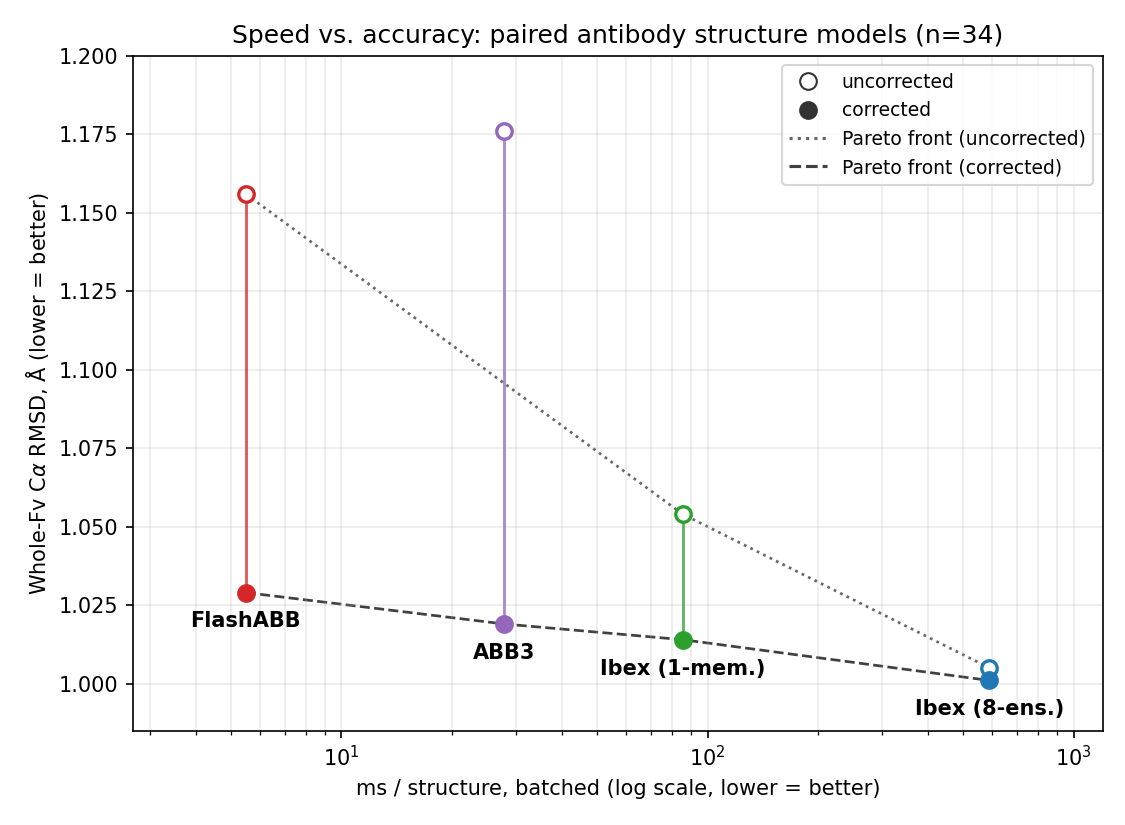}
  \caption{GPU inference speed vs. whole-Fv accuracy for the paired antibody models which comprise the Pareto front of speed and accuracy (FlashABB, ABB3, and Ibex).
  Dashed/dotted lines trace the Pareto front (no model both faster and more accurate) before and after correction.
  FlashABB and ABB3 are both single-member models, and would presumably derive a similar benefit from taking an ensemble over 8 models.
  }
  \label{fig:pareto-headline}
\end{figure}

As shown in the previous section, the corrected version of ABB3 was substantially more accurate than the uncorrected default.
We also tested the effect on FlashABB, Ibex, ESMFold, and OpenFold.
ESMFold and OpenFold were both trained for monomer structure prediction and so had relatively weak performance when evaluated on antibodies using the gap trick \citep{mirdita_colabfold_2022, evans_protein_2022}.
The full results are shown in Appendix \ref{app:full_antibody_results} alongside ABodyBuilder2.
We found that three models contributed to the Pareto front of speed and accuracy: FlashABB, ABB3, and Ibex.
The correction results for these three models are shown in Figure \ref{fig:pareto-headline}.
The correction substantially closes the accuracy gap between the models, including a full-Fv RMSD improvement of 10.9\% for FlashABB and 13.3\% for ABB3.
This enables FlashABB to achieve an RMSD within $\sim$0.02Å of Ibex while operating $\sim$100x faster.

We also saw near-uniform improvements in CDRH3 accuracy, with the exception of Ibex.
CDRH3 accuracy is incredibly noisy due to large structural variations and small sample size.
As a check on whether the negative ensemble CDR-H3 result above is a
genuine correction/ensembling interaction or a small-sample artifact of
the 34-structure subset, Appendix \ref{app:ibex-train100} repeats the same
comparison on an independent, larger sample of 100 structures from
Ibex's own training set and sees an improvement over CDRH3.

\subsection{Improved nanobody structure prediction}
\label{sec:nanobody}

Nanobodies (or VHHs) are an increasingly popular class of therapeutic due to their small size which allows them to access tighter epitopes and makes them easier to manufacture and design \citep{jovcevska_therapeutic_2020}.
In contrast to regular antibodies which contain two (paired) chains, nanobodies are comprised of only a single chain.
We tested the benefit of our correction on popular nanobody structure prediction tools NanoBodyBuilder2 (NBB2), Ibex, and NbForge.
We also tested the nanobody structure performance of general protein structure predictors ESMFold and OpenFold.
All models were tested on the test set of the SAbDab2 release \citep{capel_sabdab2_2026}.
Note that this contains substantial overlap with the training set of some of these models and so absolute performance should be interpreted with caution.
However, as shown in Table \ref{tab:nanobody}, we saw an improvement for whole-domain structure prediction in all models under our correction.
We also saw an improvement in CDRH3 performance in all models except for NbForge.
As with the Ibex paired antibody results from Section \ref{sec:headline}, the apparent NbForge performance hit is likely due to noise in CDRH3 structure prediction and does not hold up under different random subsets of SAbDab2-nano.
Table \ref{tab:nanobody} uses no recycling for ESMFold and OpenFold, for a consistent single-forward-pass comparison across every model in that table; Appendix \ref{app:nanobody-recycle3} repeats the same comparison with both models' own default recycling enabled.

\begin{table}[h]
\centering
\caption{Nanobody (VHH) full-backbone RMSD, uncorrected vs.\ both-sites
corrected (both-sites+plm\_in for Ibex), 100 real held-out VHH
structures.
NbForge is shown both correcting its usual two sites (IPA,
transition\_s) and correcting all three (adding transition\_z, on the pair
representation).
Correcting the third site slightly hurts performance, although a cross-validated sweep of $q$ found that $q=0.94$ recovered performance (not shown).
$^*$The precise training details for ESMFold, Ibex, and NbForge are not publicly available, so it is possible they were finetuned with a lower dropout rate.
$^\dagger$NBB2's (NanoBodyBuilder2) structure module (same public
\texttt{ImmuneBuilder} package as ABB2) finetunes with no dropout and so cannot be corrected.}
\label{tab:nanobody}
\begin{tabular}{llrrrr}
\toprule
Model & Condition & Full VHH & Impr. & CDR-H3 & Impr. \\
\midrule
NBB2 (1 member)$^\dagger$
  & -- & 1.241 & -- & 2.374 & -- \\
\addlinespace
NBB2 (4-member ensemble)$^\dagger$
  & -- & 1.189 & -- & 2.202 & -- \\
\addlinespace
\multirow{2}{*}{Ibex (1 member)$^*$}
  & Raw & 1.123 & --     & 2.057 & --     \\
  & +DLC   & 1.101 & +2.0\% & 1.998 & +2.9\% \\
\addlinespace
\multirow{2}{*}{Ibex (8-member ensemble)$^*$}
  & Raw & 1.099 & --     & 2.011 & --     \\
  & +DLC   & 1.097 & +0.2\% & 2.000 & +0.6\% \\
\addlinespace
\multirow{3}{*}{NbForge}
  & Raw         & 1.108 & --     & 1.902 & --     \\
  & +DLC (2 sites) & 1.072 & +3.3\% & 1.912 & $-0.5\%$ \\
  & +DLC (3 sites) & 1.090 & +1.7\% & 1.930 & $-1.5\%$ \\
\addlinespace
\multirow{2}{*}{ESMFold (no recycling)$^*$}
  & Raw & 2.080 & --     & 5.014 & --     \\
  & +DLC   & 2.069 & +0.5\% & 4.981 & +0.6\% \\
\addlinespace
\multirow{2}{*}{OpenFold (no recycling)}
  & Raw & 2.085 & --     & 4.797 & --     \\
  & +DLC   & 2.068 & +0.8\% & 4.741 & +1.2\% \\
\bottomrule
\end{tabular}
\end{table}

\subsection{More reliable confidence prediction}
\label{sec:plddt-cross-model}

Many antibody structure prediction models include a pLDDT head which can be used to estimate the error in single predictions \citep{kenlay_abodybuilder3_2024}.
Unlike in ESMFold/AlphaFold2, the pLDDT head for most antibody structure predictors is conditioned on the single representation (which suffers from the dropout+LN bias).
We tested whether our correction improved models' own estimates of the uncertainty of their predictions.
The pLDDT heads are trained to predict per-residue LDDT which we averaged over the entire sequence and over only the CDRH3 residues.
The results are shown in Table \ref{tab:plddt}.
Adding our correction substantially improved ABB3 and FlashABB pLDDT accuracy.
The Spearman correlation was nearly 2x over the entire sequence for FlashABB, in line with a much better calibrated signal.
We also saw a strong improvement in the whole-domain Spearman correlation for NbForge, although only a modest improvement in the other measures.
Both Ibex models saw little to no improvement in pLDDT after applying the correction.
The details of Ibex training are not public, so it is possible that these models were finetuned with zero dropout, and/or that the pLDDT head was trained using frozen eval weights of the main model.

\begin{table}[h]
\centering
\caption{Our correction improves models' predicted uncertainty (pLDDT) accuracy. The mean absolute error (MAE) was reduced and the Pearson and Spearman correlations were improved for all models except Ibex. It is possible that the Ibex pLDDT head was trained with frozen structure prediction weights and no dropout.}
\label{tab:plddt}
\begin{tabular}{llrrrrrr}

\toprule
& & \multicolumn{3}{c}{Variable region} & \multicolumn{3}{c}{CDR-H3} \\
\cmidrule(lr){3-5} \cmidrule(lr){6-8}
Model & Condition & MAE & Pearson & Spearman & MAE & Pearson & Spearman \\
\midrule
\multirow{2}{*}{ABB3}
  & Raw & 4.233 & $-0.648$ & $-0.265$ & 12.525 & $-0.670$ & $-0.787$ \\
  & +DLC   & 3.291 & $-0.681$ & $-0.425$ & 10.400 & $-0.664$ & $-0.788$ \\
\addlinespace
\multirow{2}{*}{FlashABB}
  & Raw & 4.641 & $-0.625$ & $-0.199$ & 11.002 & $-0.680$ & $-0.772$ \\
  & +DLC   & 3.579 & $-0.680$ & $-0.396$ & 10.035 & $-0.685$ & $-0.774$ \\
\addlinespace
\multirow{2}{*}{Ibex (paired)$^*$}
  & Raw & 3.271 & $-0.647$ & $-0.294$ & 11.997 & $-0.630$ & $-0.759$ \\
  & +DLC   & 3.287 & $-0.651$ & $-0.372$ & 11.820 & $-0.616$ & $-0.745$ \\
\addlinespace
\multirow{2}{*}{Ibex (nano)$^*$}
  & Raw & 5.613 & $-0.598$ & $-0.493$ & 8.674 & $-0.689$ & $-0.673$ \\
  & +DLC   & 5.715 & $-0.611$ & $-0.493$ & 8.896 & $-0.705$ & $-0.646$ \\
\addlinespace
\multirow{2}{*}{NbForge}
  & Raw & 6.018 & $-0.520$ & $-0.410$ & 8.811 & $-0.599$ & $-0.668$ \\
  & +DLC   & 5.964 & $-0.560$ & $-0.556$ & 8.596 & $-0.600$ & $-0.708$ \\
\bottomrule
\end{tabular}
\end{table}

\subsection{The Genie family}
\label{sec:genie3}

Genie1 \citep{lin_generating_2023}, Genie2 \citep{lin_out_2024}, and Genie3 \citep{lin_fast_2026} are generative backbone design models based on the OpenFold architecture under a diffusion framework.
The Genie models share the same structure module architecture to the regression models above, including identical dropout$\to$LayerNorm sites.
We evaluate all three with the same protocol:
at 100 evenly spaced diffusion timesteps $t$ (the Genie3 default) we sample noise, run the denoiser, and compare the corrected vs. uncorrected noise-prediction loss, on the 100-nanobody test set.

The three models differ slightly in which dropout+LayerNorm sites exist.
All three share the structure module's 2-sites-per-block pattern, but Genie1 uses only 5 structure-net blocks, whereas Genie2 and Genie3 both use 8.
Genie3's trunk contains an additional single representation IPA/transition block.
This gives 10 corrected sites for Genie1, 16 for Genie2, and 26 for Genie3 (8 structure-net blocks $\times$ 2, plus 5 trunk blocks $\times$ 2).

The results are shown in Table \ref{tab:genie-family}.
Our correction improved performance in all models, with a particularly strong effect in Genie 1 and 2.
The correction closes most of the performance gap between Genie 2 and 3 over the nanobody set.

\begin{table}[h]
\centering
\caption{Genie1/Genie2/Genie3, LN-level correction vs.\ diffusion timestep
$t$ (mean over 100 evenly spaced $t$ values), on the 100-nanobody set.}
\label{tab:genie-family}
\begin{tabular}{lrrrr}
\toprule
& Sites & Uncorrected loss & +DLC-corrected loss & Impr. \\
\midrule
Genie1 & 10 & 1.1444 & 1.1097 & $+3.03\%$ \\
Genie2 & 16 & 0.8372 & 0.8145 & $+2.71\%$ \\
Genie3 & 26 & 0.8064 & 0.7955 & $+1.35\%$ \\
\bottomrule
\end{tabular}
\end{table}

In Figure \ref{fig:genie3-overshoot} (Appendix \ref{app:additional-robustness}), we additionally plot the magnitudes and directions of the corrected and uncorrected score vectors compared to the Monte Carlo mean and the true noise for Genie3 over two random monomers.
As with the ABB3 example, the uncorrected model consistently overshoots the Monte Carlo mean with dropout on, whereas the corrected version matches it almost exactly.
However, all models consistently undershoot the true noise which may mean that although the correction helps the model loss, it may lead to overly conservative structures in practice.
We leave further investigations into the impact on designs for future work.

\subsection{QuickBind: protein-ligand docking}
\label{sec:quickbind}

QuickBind \citep{treyde_quickbind_2024} is a protein-ligand docking model based on the OpenFold IPA implementation.
As with the previous models, it contains the dropout+LayerNorm bias in the IPA and structure module transitions.
We evaluated the impact of our correction on the RMSD of QuickBind predictions over the PoseBusters Benchmark set \citep{buttenschoen_posebusters_2023}
(428 protein-ligand complexes; 427 after excluding one receptor over 2000 residues).

Compared to antibody structure prediction, docking models are more likely to predict structures in the wrong mode entirely.
In these cases, a more faithful realization of the model's underlying belief of the pose does not necessarily translate to improved RMSD.
To isolate the two factors (epitope selection vs. placement), we show the RMSD improvement over the whole set and over only those predictions which were sufficiently close to the ground truth in the uncorrected prediction.
The results are shown in Table \ref{tab:quickbind}.

\begin{table}[h]
\centering
\caption{QuickBind ligand-atom RMSD, uncorrected vs. corrected, on the
PoseBusters Benchmark set, restricted to structures where the RMSD of the uncorrected prediction is within a cutoff. ``Improved'' is the fraction of structures where the corrected prediction has lower RMSD than uncorrected. $t$ is
the paired $t$-statistic on the per-structure corrected$-$uncorrected
difference ($|t|\gtrsim1.96$ is significant at $p<0.05$).}
\label{tab:quickbind}
\begin{tabular}{lrrrrrr}
\toprule
Uncorrected RMSD threshold & $n$ & Raw & +DLC & Impr. & Improved & $t$ \\
\midrule
$<2$\,\AA        &  70 & 1.332 & 1.313 & +1.40\% &  46/70  & $-1.68$ \\
$<3$\,\AA        & 120 & 1.815 & 1.787 & +1.54\% &  76/120 & $-3.08$ \\
$<5$\,\AA        & 204 & 2.668 & 2.655 & +0.50\% & 119/204 & $-1.19$ \\
None (full set) & 427 & 9.384 & 9.343 & +0.43\% & 240/427 & $-1.07$ \\
\bottomrule
\end{tabular}
\end{table}

Over the full benchmark the effect is small and not statistically
significant ($t=-1.07$).
However, when restricted to structures which were already close to correct without the correction, the effect becomes much more consistent.
At the $<3$ Å cutoff the improvement is significant ($t=-3.08$, $p\approx0.002$) with a 63/37 skew toward improvement.
The $<2$ Å threshold is a very strong filter for positive uncorrected results, which may be artificially excluding predictions for which the correction would help.

\section{Discussion}

In this work we introduce DLC, a computationally negligible fix to a systematic bias in neural networks containing dropout layers directly upstream of LayerNorms.
The correction is computable in closed form and reliably improves performance across all 10 models tested.
In some models, such as ESMFold, the effect is small but consistent.
In others, such as ABB3, the effect is large enough to meaningfully expand the Pareto front for antibody structure prediction.

Why does the bias seem to affect some models more than others?
One hypothesis is that some models learn a more robust capacity to correct in the later layers.
For instance, ESMFold and AlphaFold2/OpenFold have shared weights for the layers in the structure module.
This means that if they overshoot after the first layer, they can likely close the distance in future layers.
This is in contrast to antibody structure prediction models like ABB3 which learn separate weights in each layer.
This may mean that these models end up learning to take single large jumps in the early layers and use the later layers for small structural tweaks.
If there is bias in the large early jumps, this may be unrecoverable later on.

One of the questions this work allows us to answer is why the relatively straightforward combination of dropout+LayerNorm is so rare in frontier models.
The bias induced by this pattern means that, paradoxically, higher dropout rates actually degrade generalization performance.
The previous solution has been to minimize the variance gap by either reducing the dropout rate or introducing a residual connection.
The residual pattern is nearly ubiquitous in modern machine learning: it is contained in GNNs like ProteinMPNN \citep{dauparas_robust_2022}, the pair module of AlphaFold2/3 \citep{jumper_highly_2021, abramson_accurate_2024}, and almost all Transformer-based language models \citep{vaswani_attention_2017, touvron_llama_2023}.
We ran a small number of experiments on small models with residual connections and found that, with a dropout rate of 10\% and typical variance ratios between the layer representations and the residual stream, this strategy does seem to practically solve the expectation gap.
However, it may be that larger models are more sensitive to the accumulated error or that dropout rates exceeding 10\% push the expectation gap far enough to impact generalization performance.
DLC offers a fairer assessment of the impact of higher dropout rates which may improve generalization in the training of frontier models.

Additionally, there are some models which show that there are benefits of dropout+LayerNorm.
CogView demonstrated that separating normalization from the residual stream (thereby introducing a dropout+LayerNorm) improved numerical stability of their vision Transformer \citep{ding_cogview_2021}.
It is likely that CogView would be improved by implementing DLC.
Our work unlocks the ability to experiment with models that include this motif, without the (previously) inexplicable validation penalty they normally incur.
It is possible that this will lead to developments in future models for protein structure prediction/design, image processing, natural language, and more.

\clearpage

\bibliographystyle{conference}
\bibliography{references}

\clearpage

\appendix
\section{Full ($\mu \neq 0$) derivation}
\label{app:general-derivation}

This appendix gives the full derivation of the general correction
$c(\mu,\sigma) = \sqrt{q/(1+p(\mu/\sigma)^2)}$ stated in
Section \ref{sec:simple-derivation} without the zero-mean simplifying
assumption, and its damped extension to residual-bypass sites. Notation
follows Section \ref{sec:theory}: $x \in \mathbb{R}^d$ is the pre-dropout
activation, $\mu := \mathrm{mean}_k(x_k)$, $\sigma^2 := \mathrm{var}_k(x_k)$,
$z_k = m_k x_k / q$ with $m_k \sim \mathrm{Bernoulli}(q)$ i.i.d., $p=1-q$.
All approximations drop $O(1/d)$ terms, which vanish as the channel
dimension $d$ grows and are already negligible at the channel widths
used throughout this paper.

\subsection{Exact per-channel moments of $z$}
Since $\mathbb{E}[m_k] = q$ and $\mathrm{Var}(m_k) = pq$,
\begin{equation}
\mathbb{E}[z_k] = x_k, \qquad
\mathrm{Var}(z_k) = \frac{\mathrm{Var}(m_k)}{q^2}\, x_k^2 = \frac{p}{q}\, x_k^2 ,
\end{equation}
and $z_k, z_j$ are independent for $k \neq j$ (the masks are drawn
independently per channel).

\subsection{Mean and variance of the LayerNorm statistics}
Write $M := \mathrm{mean}_k(z_k) = \frac{1}{d}\sum_k z_k$. Linearity gives
$\mathbb{E}[M] = \frac{1}{d}\sum_k x_k = \mu$ exactly, for any $\mu$ (this
does not require the zero-mean assumption). For the shared variance
$V := \mathrm{mean}_k(z_k^2) - M^2$, first
\begin{equation}
\mathbb{E}\!\left[\frac{1}{d}\sum_k z_k^2\right]
= \frac{1}{d}\sum_k \left(\mathrm{Var}(z_k) + \mathbb{E}[z_k]^2\right)
= \frac{1}{d}\sum_k \left(\frac{p}{q} + 1\right) x_k^2
= \frac{\sigma^2+\mu^2}{q} ,
\end{equation}
using $\frac{1}{d}\sum_k x_k^2 = \sigma^2+\mu^2$ and $1+p/q = 1/q$. Since
$\mathrm{Var}(M) = \frac{1}{d^2}\sum_k \mathrm{Var}(z_k) = \frac{p(\sigma^2+\mu^2)}{qd} = O(1/d)$,
$\mathbb{E}[M^2] = \mathrm{Var}(M) + \mu^2 \approx \mu^2$, and therefore
\begin{equation}
\label{eq:app-var-z}
\mathbb{E}[V] \;\approx\; \frac{\sigma^2+\mu^2}{q} - \mu^2 \;=\; \frac{\sigma^2 + p\mu^2}{q} ,
\end{equation}
which is exactly the shared-variance expectation quoted (for the
zero-mean case only) in Section \ref{sec:simple-derivation} -- here it holds
for general $\mu$.

\subsection{The general correction}
Treating $V$ as concentrated near its mean (the same leading-order
decoupling of numerator and denominator used in
Section \ref{sec:simple-derivation}),
\begin{equation}
\mathbb{E}[\mathrm{norm}(z)_k] \;\approx\; \frac{\mathbb{E}[z_k]-\mathbb{E}[M]}{\sqrt{\mathbb{E}[V]}}
\;=\; \frac{x_k-\mu}{\sqrt{(\sigma^2+p\mu^2)/q}}
\;=\; \sqrt{\frac{q}{1+p(\mu/\sigma)^2}}\cdot \mathrm{norm}(x)_k .
\end{equation}
Defining
\begin{equation}
\label{eq:app-general-c}
c(\mu,\sigma) \;:=\; \sqrt{\frac{q}{1+p(\mu/\sigma)^2}} ,
\end{equation}
the same $\beta$-preserving algebra as Eq.~\ref{eq:simple-correction}
applies unchanged, with $c(\mu,\sigma)$ in place of $\sqrt{q}$:
\begin{equation}
\widehat{\mathrm{LN}}(x) \;=\; c(\mu,\sigma)\cdot \mathrm{LN}_{\mathrm{full}}(x) \;+\; \left(1-c(\mu,\sigma)\right)\beta .
\end{equation}
Setting $\mu=0$ recovers $c=\sqrt{q}$
(Eq.~\ref{eq:simple-correction}) as the special case, as claimed in
Section \ref{sec:simple-derivation}. This is the formula, computed live from
each residue's own pre-dropout $(\mu,\sigma)$, used at every undamped
correction site in this paper.

\subsection{Damped variant for residual-bypass sites}
\label{app:damped-derivation}
Some sites do not feed dropout's output directly into LayerNorm; instead
a bypass term is added first, $y = h + z$, where $h$ is untouched by
this dropout call and $z_k = m_k v_k/q$ is the dropped-out contribution
(e.g.\ ProteinMPNN-style $h+\mathrm{dropout}(v)$, or Genie3's
\texttt{PairTransition} input, which sums an undropped running pair
representation with two independently-dropped triangle-update terms --
Section \ref{sec:genie3}). Let $x := h+v$ denote the (deterministic,
no-dropout) combined signal, $\mu:=\mathrm{mean}_k(x_k)$,
$\sigma^2:=\mathrm{var}_k(x_k)$. Repeating the moment calculation with
$h_k$ held fixed (it carries no dropout randomness at this site):
\begin{equation}
\mathbb{E}\!\left[\frac{1}{d}\sum_k y_k^2\right]
= \frac{1}{d}\sum_k\!\left[(h_k+v_k)^2 + \frac{p}{q}v_k^2\right]
= (\sigma^2+\mu^2) + \frac{p}{q}\overline{v^2} ,
\end{equation}
where $\overline{v^2} := \frac{1}{d}\sum_k v_k^2$ is the mean-square of
the dropped-out term alone. As before $\mathbb{E}[\mathrm{mean}_k(y)^2]\approx\mu^2$, so
\begin{equation}
\mathbb{E}[V_y] \;\approx\; \sigma^2 + \frac{p}{q}\,\overline{v^2}
\;=\; \sigma^2\left(1+\frac{p}{q}\rho^2\right), \qquad
\rho^2 \;:=\; \frac{\overline{v^2}}{\sigma^2} ,
\end{equation}
and following the same steps as above,
\begin{equation}
\label{eq:app-damped-c}
\mathbb{E}[\mathrm{norm}(y)_k] \;\approx\; \sqrt{\frac{q}{q+p\rho^2}}\cdot\mathrm{norm}(x)_k ,
\qquad
c_{\mathrm{damped}} \;:=\; \sqrt{\frac{q}{q+p\rho^2}} .
\end{equation}
$\rho^2$ is the ratio of the combined signal's mean-square that comes
from the dropped-out term $v$ relative to the full signal's variance; it
is $0$ when the bypass totally dominates ($v\to 0$, so $c_{\mathrm{damped}}\to 1$,
no correction needed) and grows as $v$'s contribution grows.
Eq.~\ref{eq:app-damped-c} is the general case:
setting $h=0$ (no bypass, $v=x$, so $\overline{v^2}=\sigma^2+\mu^2$ and
$\rho^2 = 1+(\mu/\sigma)^2$) gives
$c_{\mathrm{damped}} = \sqrt{q/(q+p+p(\mu/\sigma)^2)} = \sqrt{q/(1+p(\mu/\sigma)^2)} = c(\mu,\sigma)$,
exactly recovering Eq.~\ref{eq:app-general-c} -- the undamped correction
used everywhere else in this paper is the $h=0$ special case of the same
formula.

\section{Extension to other normalization schemes}
\label{app:other-norms}

The dropout-induced shared-denominator mechanism analyzed in Section \ref{sec:theory} and Appendix \ref{app:general-derivation} is not specific to LayerNorm.
We derive the equivalent correction for two other widely used normalization schemes: RMSNorm \citep{zhang_root_2019}, which shares LayerNorm's per-example normalization but omits mean-centering, and BatchNorm \citep{ioffe_batch_2015}, which normalizes using statistics computed across the batch (and, at eval time, a moving average accumulated during training) rather than per example. Neither is used by any model tested in this paper, so both corrections are theoretical
extensions, not empirically validated here.

\subsection{RMSNorm}
\label{app:rmsnorm}

RMSNorm normalizes by the root-mean-square rather than the standard deviation, with no mean subtraction and (conventionally) no shift parameter:
\begin{equation}
\mathrm{RMSNorm}(x)_k = \gamma_k \cdot \frac{x_k}{\mathrm{rms}(x)}, \qquad
\mathrm{rms}(x) := \sqrt{\mathrm{mean}_k(x_k^2) + \epsilon} .
\end{equation}
With $z_k = m_k x_k/q$ as in Section \ref{sec:theory}, RMSNorm's denominator
uses the raw (uncentered) second moment, which was already computed
exactly in Appendix \ref{app:general-derivation} (the numerator of
Eq.~\ref{eq:app-var-z}, before the $-\mu^2$ centering term is subtracted):
\begin{equation}
\mathbb{E}\!\left[\mathrm{mean}_k(z_k^2)\right] = \frac{\sigma^2+\mu^2}{q} = \frac{S^2}{q}, \qquad S^2 := \mathrm{mean}_k(x_k^2) .
\end{equation}
Using the same concentration approximation as
Section \ref{sec:simple-derivation},
\begin{equation}
\mathbb{E}[\mathrm{RMSnorm}(z)_k] \;\approx\; \frac{x_k}{\sqrt{S^2/q}} \;=\; \sqrt{q}\cdot\mathrm{RMSnorm}(x)_k ,
\end{equation}
so $c_{\mathrm{RMS}} = \sqrt{q}$ -- the same constant as the zero-mean
LayerNorm case (Eq.~\ref{eq:simple-correction}), except here it is
\emph{exact for every $\mu$}, not just $\mu\approx0$. LayerNorm's
$\mu$-dependence (Eq.~\ref{eq:app-general-c}) comes entirely from
subtracting a non-inflated $\mathbb{E}[M]^2\approx\mu^2$ from the inflated
raw second moment $(\sigma^2+\mu^2)/q$; RMSNorm never performs that
subtraction, so its correction collapses to the single global constant
$\sqrt{q}$ regardless of $\mu$. The corrected estimator is simply
$\widehat{\mathrm{RMSNorm}}(x) = \sqrt{q}\cdot\mathrm{RMSNorm}(x)$.

\subsection{BatchNorm}
\label{app:batchnorm}

BatchNorm normalizes each channel using statistics computed across the
batch dimension rather than across channels within one example, and at
eval time replaces these with a fixed moving average accumulated during
training:
\begin{equation}
\mathrm{BN}(x)_k = \gamma_k\cdot\frac{x_k - \mathrm{running\_mean}_k}{\sqrt{\mathrm{running\_var}_k+\epsilon}} + \beta_k .
\end{equation}
Unlike LayerNorm, this is a \emph{fixed affine map} at eval time (the normalizing statistics do not depend on the current input) so there is no per-example shared-denominator randomness to analyze. Instead, the question is whether $\mathrm{running\_mean}_k,\mathrm{running\_var}_k$, accumulated from dropout-corrupted training batches, correctly reflect the dropout-free statistics that a test-time (dropout-off) input actually has.

Let $\mu_k,\sigma_k^2$ denote channel $k$'s true population mean and variance over the data distribution (no dropout).
During training, each batch element's pre-BN activation is itself dropped out, $z_k^{(b)}=m_k^{(b)}x_k^{(b)}/q$; repeating the moment calculation of Appendix \ref{app:general-derivation}, but marginalizing over the data distribution (index $b$, across the batch) instead of over channels (index $k$), gives the same expression with the roles of $\mu,\sigma$ swapped from a per-example, cross-channel quantity to a per-channel, cross-example one:
\begin{equation}
\label{eq:app-bn-runningvar}
\mathrm{running\_mean}_k \to \mu_k, \qquad
\mathrm{running\_var}_k \to \frac{\sigma_k^2+p\mu_k^2}{q} .
\end{equation}
At test time, dropout is off, so BatchNorm's fixed affine map is applied
directly to $x_k$ (not to any dropped-out quantity), and the
\emph{correct}, uninflated denominator would be $\sigma_k$, not
$\sqrt{\mathrm{running\_var}_k}$. The ratio between the two is
\begin{equation}
c(\mu_k,\sigma_k) \;=\; \frac{\sigma_k}{\sqrt{\mathrm{running\_var}_k}} \;=\; \sqrt{\frac{q\,\sigma_k^2}{\sigma_k^2+p\mu_k^2}} \;=\; \sqrt{\frac{q}{1+p(\mu_k/\sigma_k)^2}} ,
\end{equation}
exactly the same functional form as LayerNorm's general correction (Eq.~\ref{eq:app-general-c}), but with $\mu,\sigma$ now the per-channel statistics across the training population rather than the cross-channel statistics of a single example. This matches (once translated to our notation) the variance-shift ratio $\Delta(q)=\mathrm{Var}_{\mathrm{test}}(X)/\mathrm{Var}_{\mathrm{train}}(X)$ derived independently by \citet{li_understanding_2018} for Dropout-directly-before-BatchNorm ($\Delta(q)=c(\mu,\sigma)^2$): their result establishes that this variance mismatch degrades accuracy in convolutional networks, but does not give a closed-form correction; $c(\mu_k,\sigma_k)$ above is that correction.

Because $\mathrm{running\_mean}_k=\mu_k$ exactly (dropout is unbiased), Eq.~\ref{eq:app-bn-runningvar} can be inverted to express the correction purely in terms of BatchNorm's own two buffers, with no separate estimate of $\mu_k,\sigma_k$ required:
\begin{equation}
\label{eq:app-bn-correction}
c_k \;=\; \sqrt{q - p\cdot\frac{\mathrm{running\_mean}_k^2}{\mathrm{running\_var}_k}} ,
\end{equation}
and the same $\beta$-preserving correction as Eq.~\ref{eq:simple-correction} applies with $c_k$ in place of $\sqrt{q}$.
Unlike the LayerNorm/RMSNorm corrections, which need a live hook on every forward pass's pre-dropout activation (since $\mu,\sigma$ vary per example), Eq.~\ref{eq:app-bn-correction} is a \emph{one-time} transform of a trained model's frozen buffers -- $c_k$ can be computed once, after training, and folded directly into $\mathrm{running\_var}_k$ (or $\gamma_k$), at zero added inference cost.

\section{Further pLDDT calibration}
\label{app:plddt}

ABB3-style structure models feed the final block's corrected
output directly into a second LayerNorm.
The second LayerNorm removes the rescaling, which means the pLDDT head receives a biased representation.
This site is structurally distinct from every correction site used elsewhere
in this paper: the randomness it inherits is not a fresh elementwise Bernoulli mask, but the residual variance of $z_1 = \mathrm{LN}_1(\mathrm{dropout}(x))$
left over \emph{after} matching its mean via $\hat{c}_1$ -- i.e.\ a second
nonlinear (shared-denominator) operation stacked on top of an
already-corrected quantity.

\subsection{An effective correction can be derived and validated}
Applying the moment-calculation technique of
Section \ref{app:damped-derivation} to this compound case (deriving
$\mathrm{Var}(\mathrm{norm}(z_1)_k)$ from first principles rather than
assuming the raw-dropout template applies) gives a damped-shaped
correction $c_2 = \sqrt{q/(q+p\rho^2)}$ with
$\rho^2 = c_1^2 \cdot S / (\sigma_x^2 \cdot \mathrm{Var}_k(\hat{c}_1))$,
$S := \mathrm{mean}_k((\mathrm{weight}_1)_k^2 x_k^2)$ -- computable live
from quantities already available at this site (the pre-dropout $x$,
LN$_1$'s own weight, and $c_1$). Two independent checks confirm this
derivation is correct: (1) isolating \emph{only} the last transition
block's dropout (every other stochastic site held at eval, so $x$ is
exactly the fixed quantity the derivation assumes) and comparing against
a 100-sample Monte Carlo estimate of the true $\mathbb{E}[\mathrm{LN}_2(z_1)]$,
the $\rho^2$-corrected estimate cuts the mean gap to ground truth roughly
in half versus a mean-only ($c_1$-only) correction (0.255 vs.\ 0.507,
$n{=}2273$ residues); (2) a sweep over a multiplier $k$ on $p$ inside the
$c_2$ formula confirms $k{=}1$ (the as-derived formula) minimizes this
gap, ruling out a missing constant factor.

\subsection{The correct estimate is poorly calibrated}

Despite (1)-(2), applying this correction -- or the model's own real
$p{=}0.1$, or an empirically curve-fit constant $p$ -- to
\texttt{pLDDT\_head.layer\_norm} \emph{degrades} pLDDT calibration
(mean absolute error between predicted pLDDT and the true lDDT-C$\alpha$
of that same predicted structure), consistently on both the 34-structure
test subset and an independent random 100-structure sample of ABB3's own
train split (ruling out an out-of-distribution explanation):
\begin{center}
\begin{tabular}{lrr}
\toprule
Correction & MAE (test) & MAE (train, 100) \\
\midrule
Uncorrected (status quo) & 3.315 & 3.679 \\
Model's real $p{=}0.1$ (naive reuse) & 3.412 ($-2.9\%$) & 3.763 ($-2.3\%$) \\
Empirically fit constant $p{=}0.157$ & 3.481 ($-5.0\%$) & 3.822 ($-3.9\%$) \\
Derived $\rho^2$ (per-residue, live) & 3.396 ($-2.4\%$) & 3.751 ($-2.0\%$) \\
\bottomrule
\end{tabular}
\end{center}
The ranking (derived $\rho^2$ least harmful, flat empirical constant most
harmful) tracks how theoretically grounded each variant is, so this isn't
a broken derivation -- it's a correct estimate of the wrong target.
Sweeping a free scalar $c_2$ directly against calibration (rather than
against the true $\mathbb{E}[\mathrm{LN}_2(z_1)]$) finds a real, reproducible
optimum at $c_2 \approx 1.75$ -- i.e.\ \emph{amplifying} away from
$\mathrm{bias}_2$, the opposite direction from every theoretically-motivated
correction above -- giving a genuine $\sim\!7\%$ MAE improvement,
identically on both test ($3.315\to 3.079$) and train ($3.679\to 3.422$).

Correcting the structure module measurably improves accuracy but mean predicted pLDDT is essentially flat.
pLDDT's \emph{ranking} of residues by expected accuracy is not broken.
Spearman correlation with true C$\alpha$ error is actually strongest at
$c_2{=}1$ ($-0.425$) and the derived $\rho^2$ correction ($-0.431$), both better than uncorrected ($-0.265$), but its \emph{absolute scale} is stuck calibrated to the noisier, pre-correction accuracy regime pLDDT was
trained against, since dropout was always on during training. The empirically-optimal $c_2{=}1.75$ partially repairs this absolute mismatch (mean pLDDT rises to $95.78$, appropriately higher for the now-better structures) at a real cost to discrimination (Spearman drops to $-0.384$).
A coarse single-scalar rescale can only shift the whole
distribution, not fix calibration and sharpness simultaneously the way retraining \texttt{pLDDT\_head} against corrected-model outputs presumably would.

\section{Additional robustness checks}
\label{app:additional-robustness}

\subsection{Full antibody results}
\label{app:full_antibody_results}

\begin{table}[H]
\centering
\caption{
Uncorrected vs.\ corrected RMSDs for the paired antibody models on the 34-structure test subset.
Predictions do not include MD refinement.
CDR-H3 RMSDs are computed by aligning to heavy framework residues.
Ibex is reported both as a single ensemble member (member 0 only, no ensembling) and as the full 8-member deep ensemble (closest model to mean).
ESMFold and OpenFold are run using the AlphaFold ``gap trick'' with no recycling.
$^*$We do not have access to ESMFold's or Ibex's exact training details, so it is possible they were finetuned with no dropout.
$^\dagger$ABB2 finetunes with no dropout and so cannot be corrected.}
\label{tab:headline}
\begin{tabular}{llrrrr}
\toprule
Model & Condition & Whole-Fv & Impr. & CDR-H3 & Impr. \\
\midrule
ABB2 (1 member)$^\dagger$
  & -- & 1.110 & -- & 2.968 & -- \\
\addlinespace
ABB2 (4-member ensemble)$^\dagger$
  & -- & 1.095 & -- & 2.862 & -- \\
\addlinespace
\multirow{2}{*}{ABB3}
  & Raw & 1.176 & --     & 2.861 & --    \\
  & +DLC   & 1.019 & +13.3\% & 2.741 & +4.2\% \\
\addlinespace
\multirow{2}{*}{FlashABB}
  & Raw & 1.156 & --     & 2.695 & --    \\
  & +DLC   & 1.029 & +10.9\% & 2.656 & +1.5\% \\
\addlinespace
\multirow{2}{*}{Ibex (1 member)$^*$}
  & Raw & 1.054 & --    & 2.846 & --    \\
  & +DLC   & 1.014 & +3.8\% & 2.791 & +2.0\% \\
\addlinespace
\multirow{2}{*}{Ibex (8-member ensemble)$^*$}
  & Raw & 1.005 & --     & 2.675 & --     \\
  & +DLC   & 1.001 & +0.5\% & 2.712 & $-1.4\%$ \\
\addlinespace
\multirow{2}{*}{ESMFold$^*$}
  & Raw & 1.423 & --     & 3.232 & --     \\
  & +DLC   & 1.420 & +0.2\% & 3.224 & +0.2\% \\
\addlinespace
\multirow{2}{*}{OpenFold}
  & Raw & 1.420 & --     & 3.691 & --     \\
  & +DLC   & 1.417 & +0.2\% & 3.674 & +0.5\% \\
\bottomrule
\end{tabular}
\end{table}

\subsection{Ibex on a random 100-structure training-set subsample}
\label{app:ibex-train100}
As a check on whether the negative ensemble CDR-H3 result of
Section \ref{sec:headline} is a genuine correction/ensembling interaction or a
small-sample artifact of the 34-structure test subset, this repeats the
same comparison on an independent, much larger sample: 100 structures
drawn at random (seed 42) from Ibex's own published training set
(\texttt{docs/ibex\_train.csv} in the Ibex source repository), restricted
to the 8026/8395 structures already overlapping our own training split
(so ground-truth atom positions and AHO-scheme CDR-H3 boundaries, the
latter recomputed per structure via ANARCI, did not need to be rebuilt
from scratch) \citep{dunbar_anarci_2016}.
Framework-aligned CDR-H3 and whole-Fv RMSD, same
convention as Table~\ref{tab:headline}.

\begin{table}[h]
\centering
\caption{Ibex correction impact on a random 100-structure subsample of
Ibex's own training set (not the 34-structure held-out test subset used
elsewhere in this paper), member 0 and the 8-member medoid ensemble.
$^*$We do not have access to Ibex's exact training/finetuning details
(e.g.\ whether dropout was disabled during finetuning, as in
ImmuneBuilder's convention), so we cannot rule out that this affects
how the correction interacts with this model specifically.}
\label{tab:ibex-train100}
\begin{tabular}{llrrrr}
\toprule
Model & Condition & Whole-Fv & Impr. & CDR-H3 & Impr. \\
\midrule
\multirow{3}{*}{Ibex (1 member)$^*$}
  & Raw         & 0.9070 & --     & 1.479 & --     \\
  & Both sites          & 0.8800 & +3.0\% & 1.419 & +4.1\% \\
  & Both sites + plm\_in & 0.8793 & +3.1\% & 1.411 & +4.6\% \\
\addlinespace
\multirow{3}{*}{Ibex (8-member ensemble)$^*$}
  & Raw         & 0.8926 & --     & 1.431 & --     \\
  & Both sites          & 0.8786 & +1.6\% & 1.408 & +1.6\% \\
  & Both sites + plm\_in & 0.8760 & +1.9\% & 1.388 & +3.0\% \\
\bottomrule
\end{tabular}
\end{table}

\subsection{ESMFold and OpenFold with their own default recycling}
\label{app:nanobody-recycle3}
Table~\ref{tab:nanobody} uses no-recycling for ESMFold and OpenFold, for a
consistent single-forward-pass comparison across every model in that
table. Both models also support recycling by default (3 iterations); this
table repeats the same nanobody-100 comparison with recycling enabled,
same full-backbone/CDR-H3 convention throughout.

\begin{table}[h]
\centering
\caption{ESMFold and OpenFold on the nanobody-100 set with 3 recycling
iterations (their own default), rather than the no-recycling convention
used in Table~\ref{tab:nanobody}. $^*$We do not have access to ESMFold's
exact training/finetuning details, so it is possible it was finetuned
with no dropout.}
\label{tab:nanobody-recycle3}
\begin{tabular}{llrrrr}
\toprule
Model & Condition & Whole-domain & Impr. & CDR-H3 & Impr. \\
\midrule
\multirow{2}{*}{ESMFold (3 recycles)$^*$}
  & Raw & 1.835 & --     & 4.186 & --     \\
  & +DLC   & 1.827 & +0.4\% & 4.160 & +0.6\% \\
\addlinespace
\multirow{2}{*}{OpenFold (3 recycles)}
  & Raw & 1.721 & --     & 3.654 & --     \\
  & +DLC   & 1.716 & +0.3\% & 3.639 & +0.4\% \\
\bottomrule
\end{tabular}
\end{table}

\subsection{Statistical significance of the ESMFold/OpenFold improvement}
\label{app:esmfold-openfold-significance}

Table~\ref{tab:nanobody}'s ESMFold and OpenFold improvements are modest in
absolute terms ($+0.5\%$, $+0.8\%$) compared to the antibody-specific
models, so we checked whether they are statistically distinguishable from
noise using the same paired-$t$ convention as
Table~\ref{tab:quickbind} ($|t|\gtrsim1.96 \Rightarrow p<0.05$), on the
per-structure corrected$-$uncorrected difference over the same 100
nanobody structures.

\begin{figure}[h]
  \centering
  \includegraphics[width=\textwidth]{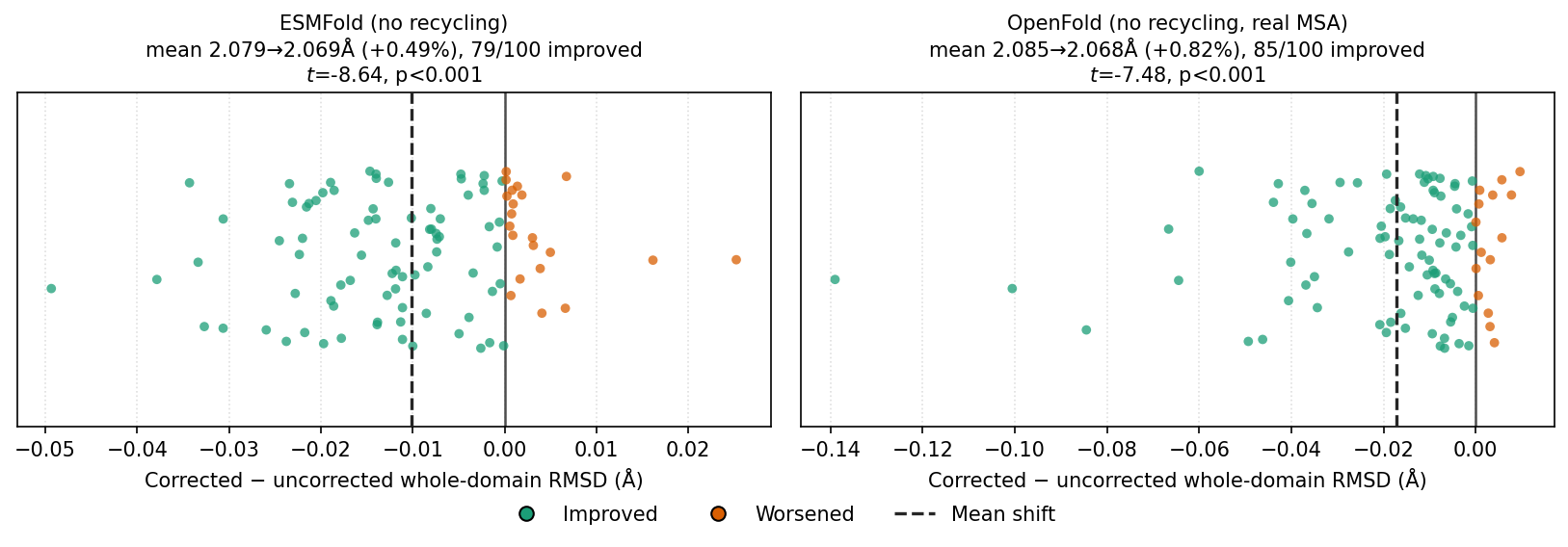}
  \caption{Per-structure whole-domain RMSD shift (corrected$-$uncorrected)
  for ESMFold and OpenFold, no-recycling, on the same 100 nanobody
  structures as Table~\ref{tab:nanobody}. Each point is one structure;
  green/orange indicate whether that structure's RMSD improved or
  worsened under correction. Both shifts are significant despite their
  small mean magnitude ($t=-8.64$, $p<0.001$ for ESMFold; $t=-7.48$,
  $p<0.001$ for OpenFold) -- the correction reliably nudges the large
  majority of structures (79/100 and 85/100) in the same direction, it is
  just a small nudge per structure for these two general-purpose,
  non-antibody-specific models.}
  \label{fig:esmfold-openfold-shift}
\end{figure}

\subsection{Calibration: sweeping the assumed $q$}
\label{sec:qsweep}

\begin{figure}[H]
  \centering
  \includegraphics[width=\textwidth]{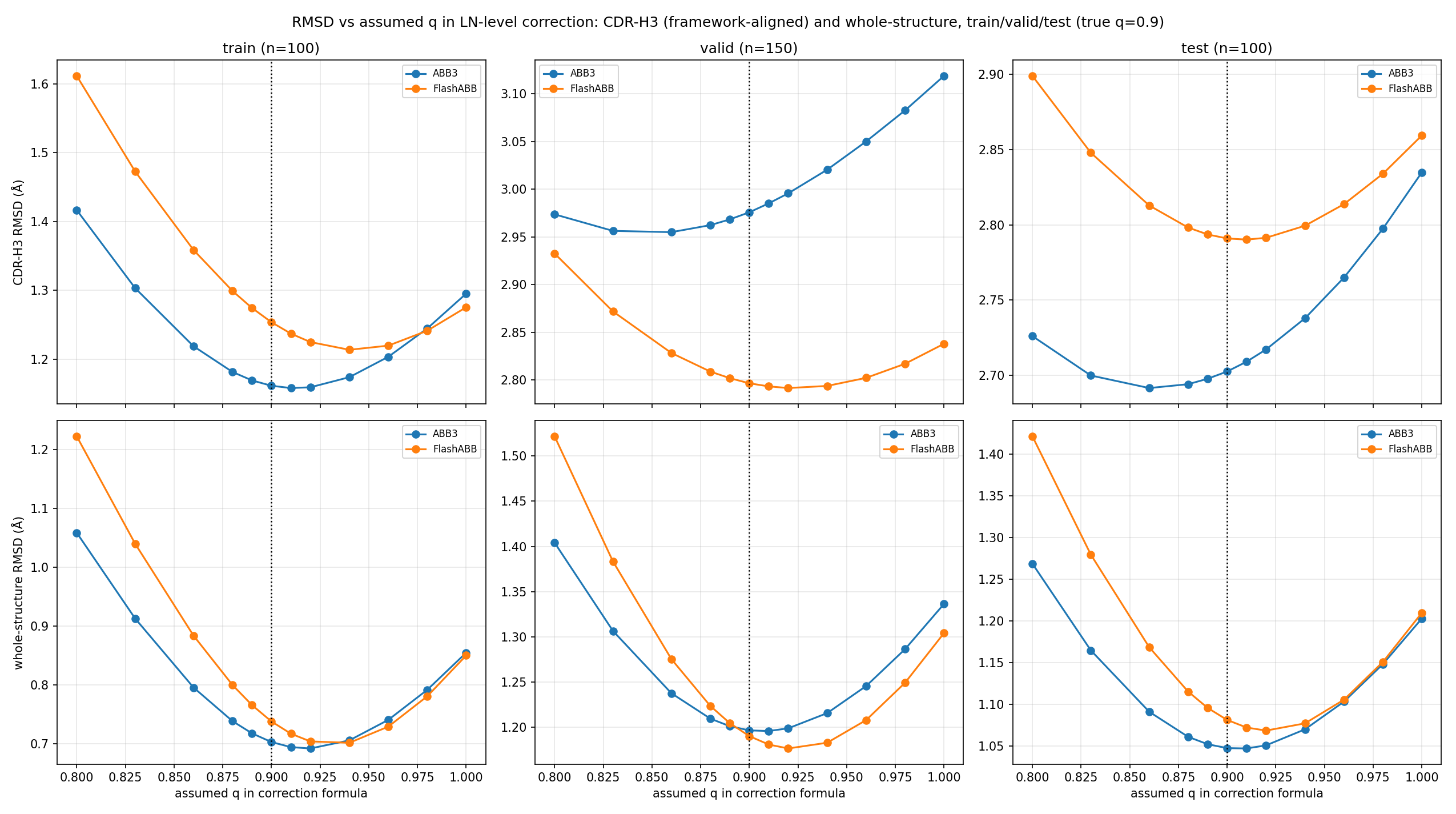}
  \caption{RMSD as a function of the assumed keep-probability $q$ used
  inside the correction formula (Eq.~\ref{eq:simple-correction}/general
  form), for CDR-H3 (framework-aligned) and whole-structure RMSD, on
  train/valid/test splits, ABB3 and FlashABB. The true configured
  $q=0.9$ is marked.}
  \label{fig:q-sweep}
\end{figure}

Throughout this work, we correct using the reported dropout rates for each model (typically 0.1).
In Figure \ref{fig:q-sweep}, we show RMSD as a function of the assumed keep rate, $q$, for ABB3 and FlashABB.
In all cases, $q=0.9$ is better than the uncorrected version $q=1$.
However, the optimal values did not always sit at exactly $q=0.9$.
It is possible that this is simply a noise artifact driven by large outliers favouring slight (over/under)shoot predictions.
Additionally, although we show that the correction fixes the overshoot problem almost exactly (Figure \ref{fig:abb3-block0}), it may be that downstream layers are sensitive to changes in variance induced by our correction.

\subsection{Genie3 noise-prediction decomposition}
\label{app:genie3-figure}

\begin{figure}[h]
  \centering
  \includegraphics[width=\textwidth]{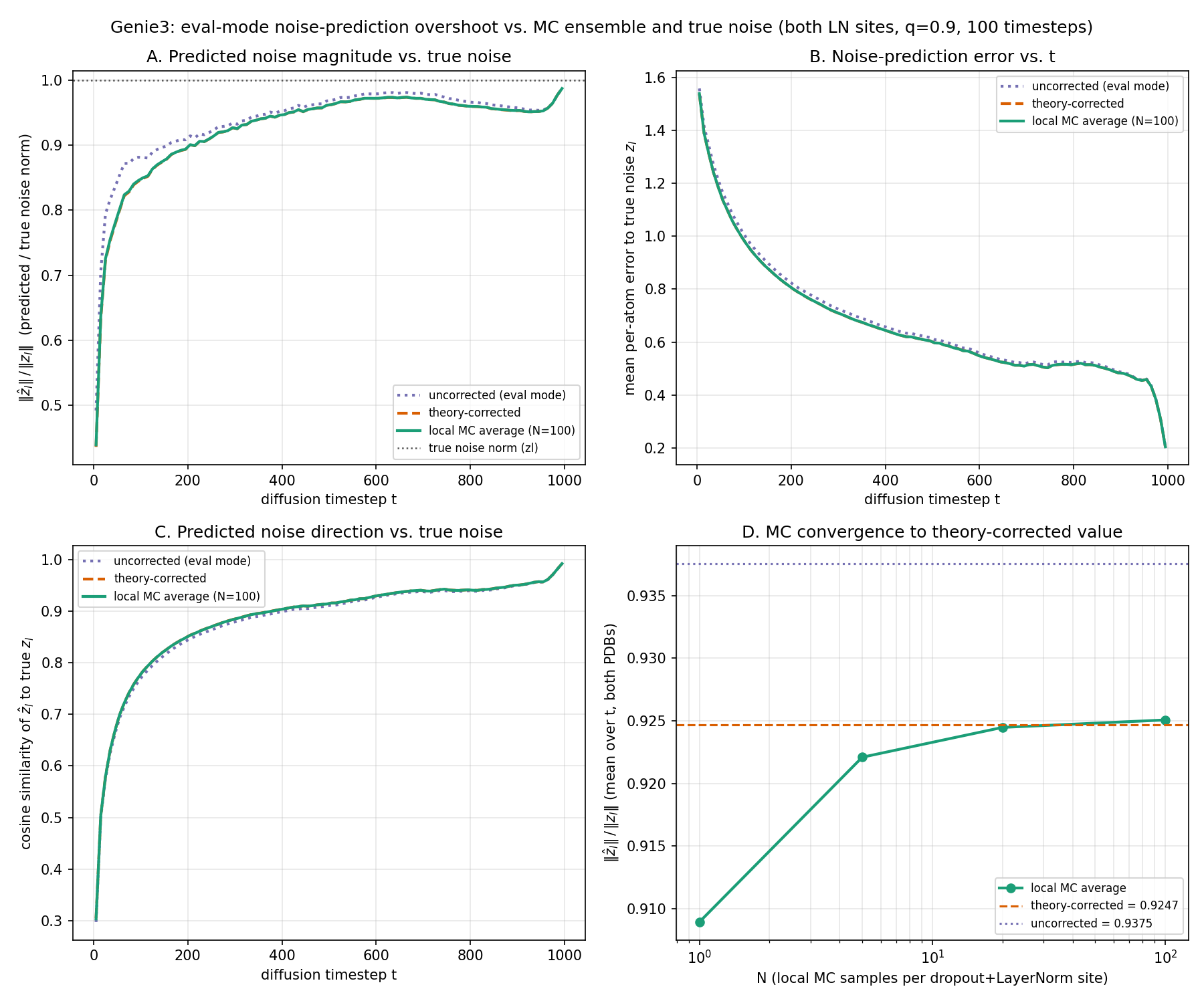}
  \caption{Genie3 noise-prediction decomposition vs.\ diffusion timestep
  $t$ (all 26 LN sites corrected -- \texttt{structure\_net} and
  \texttt{pair\_transform\_net}, $q=0.9$, 100 evenly spaced $t$ values, 2
  monomer structures $\times$ 12 noise draws per $t$). \textbf{A}:
  predicted/true noise-norm ratio -- uncorrected overshoots the
  $N{=}100$ MC ceiling at 97/100 $t$ values; all modes undershoot the
  true-noise reference line ($y=1$). \textbf{B}: per-atom error to the
  true noise vector $z_l$, decomposing Table~\ref{tab:genie-family}'s loss
  numbers. \textbf{C}: cosine similarity of predicted to true noise
  direction -- nearly identical across modes, showing the bias is a
  magnitude effect only. \textbf{D}: local MC average (mean over all $t$
  and both structures) converging onto the theory-corrected value as $N$
  grows from 1 to 100, well below the uncorrected reference.}
  \label{fig:genie3-overshoot}
\end{figure}

\end{document}